\documentclass{ws-procs9x6}
\usepackage{graphics}

\begin{document}

\newcommand{\EE}        {\ensuremath{\mathrm{e}^+\mathrm{e}^-}} 
\newcommand{\GG}        {\ensuremath{\mathrm{\gamma\gamma}}} 
\newcommand{\GSGS}      {\ensuremath{\mathrm{\gamma^{*}\gamma^{*}}}} 
\newcommand{\PP}        {\ensuremath{\mathrm{p\bar{p}}}} 
\newcommand{\LL}        {\ensuremath{\mathrm{\Lambda\overline{\Lambda}}}} 
\newcommand{\XIXI}      {\ensuremath{\mathrm{\Xi^{-}\Xi^{+}}}}
\newcommand{\SISI}        {\ensuremath{\mathrm{\Sigma^{0}\overline{\Sigma^{0}}}}}
\newcommand{\BB}        {\ensuremath{\mathrm{B\overline{B}}}}

\newcommand{\MV}        {\ensuremath{\mathrm{MeV}}}
\newcommand{\GV}        {\ensuremath{\mathrm{GeV}}}
\newcommand{\SI}        {\ensuremath{\mathrm{\sigma}}}
\newcommand{\costs}     {\ensuremath{\mathrm{|\cos\theta^{*}|}}}
\newcommand{\cost}      {\ensuremath{\mathrm{|\cos\theta|}}}
\newcommand{\mb}        {\ensuremath{\mathrm{mb}}}
\newcommand{\nb}        {\ensuremath{\mathrm{nb}}}
\newcommand{\pb}        {\ensuremath{\mathrm{pb}}}

\newcommand{\T}         {\ensuremath{\mathrm{\theta}}}
\newcommand{\LUM}       {\ensuremath{L}}
\newcommand{\NE}        {\ensuremath{N_{\rm ev}}}
\newcommand{\EDET}      {\ensuremath{\mathrm{\varepsilon}_{\rm DET}}}
\newcommand{\ETRIG}     {\ensuremath{\mathrm{\varepsilon}_{\rm TRIG}}}
\newcommand{\SQS}       {\ensuremath{\sqrt{s_{\rm ee}}}}
\newcommand{\NS}        {\ensuremath{N_{\rm SEL}}}
\newcommand{\NGEN}      {\ensuremath{N_{\rm GEN}}}
\newcommand{\D}         {\ensuremath{\mathrm{\bigtriangleup}}}
\newcommand{\PM}        {\ensuremath{\mathrm{\pm}}}
\newcommand{\PT}        {\ensuremath{p_{\perp}}}
\newcommand{\lumi}      {\ensuremath{\mathcal{L}}}
\newcommand{\cal}       {\ensuremath{\mathcal{L}}}
\newcommand{\WGG} {\ensuremath{\sf W}}

%\title{Instructions for Producing a Camera-Ready Manuscript 
%using Latex for Publication in Conference 
%Proceedings}
%Proceedings\footnote{\uppercase{T}his work is supported by etc, etc.}}

\title{Exclusive baryon-antibaryon production in $\GG$ collisions at $\EE$ colliders }

\author{T. BARILLARI}

\address{Max-Planck-Inst. f\"ur Physik, \\
Werner-Heisenberg-Institut, \\ 
F\"ohringer Ring 6,\\
D-80805 M\"unchen, Germany\\ 
E-mail: barilla@mppmu.mpg.de}

\maketitle

\abstracts{
 The exclusive production of baryon-antibaryon pairs in the collisions
 of two quasi-real photons has been studied using different detectors 
 at $\EE$ colliders.
 Results are presented for $\GG\to\PP$, $\GG\to\LL$, and $\GG\to \SISI$
 final states.   
 The cross-section measurements are compared with all the existing experimental 
 data and with the analytic calculations based on the three-quark model, on the
 quark-diquark model, and on the handbag model.}

\vspace*{-0.9cm}
\section{Introduction}
The exclusive production of baryon-antybaryon ($\BB$) pairs in the collision of
two quasi-real photons can be used to test predictions of QCD. 
At $\EE$ colliders the photons are emitted by the beam electrons\footnote{In
this paper positrons are also referred to as electrons.}
and the $\BB$ pairs are produced in the process $\EE\to\EE\GG\to\EE\BB$.

The application of QCD to exclusive photon-photon reactions
is based on the work of Brodsky and Lepage~\cite{Lepage:1980fj}. 
According to their formalism the process is factorized into a 
non-perturbative part, which is the hadronic wave function of the final 
state, and a perturbative part.
Calculations based on this 
ansatz~\cite{Farrar:1985gv,Chernyak:1984bm}
yields e.g. $\EE\to\EE\GG\to\EE\PP$ cross-sections about one 
order of magnitude smaller than the existing experimental 
results~\cite{Albrecht:1989hz,Artuso:1994xk,Hamasaki:1997cy,Abbiendi:2002bx,Achard:2003jc,Kuo:2005nr}, 
for $\PP$ centre-of-mass energies $W$ greater than $2.5\,\GV$.

Recent studies~\cite{berger:2002vc} have extended the systematic 
investigation of hard exclusive reactions within the quark-diquark model 
to photon-photon processes~\cite{Anselmino:1989gu,Kroll:1991a}.
In addition, the handbag contribution~\cite{diehl:2003} has been recently 
proposed to describe the photon-photon annihilation into baryon-antibaryon 
pairs at large momentum transfer.

In this paper, all the existing measurements of the cross-sections 
for the exclusive $\EE\to\EE \BB$ processes are presented.
In particular, results for $\GG\to\PP$, $\GG\to\LL$, and $\GG\to \SISI$
final states are reported. 
These cross-section measurements are compared 
with the analytic calculations based on the three-quark model, on the
quark-diquark model, and on the handbag model. 

\section{The $\GG\to\PP$ cross-section measurements}
\label{sec:ppbar cross-section}

The differential cross-section for the process $\EE\to\EE\PP$ is given by
\begin{equation*}
  \frac{{\rm d}^2\SI(\EE\to\EE\PP)}{{\rm d}W\,{\rm d}\costs} = 
  \frac{{\NE(W,\costs)}}{{\lumi}_{\EE}\ETRIG\,\EDET\,(W,\costs)\,\Delta W\,\Delta\costs}
\label{eq:diffcross}
\end{equation*}

\vspace*{-0.4cm}

\noindent where \NE\ is the number of events selected in each $(W,\costs)$
bin, $\ETRIG$ is the trigger efficiency, $\EDET$ is the detection 
efficiency, $\lumi_{\EE}$ is the measured integrated luminosity, and 
$\Delta W$ and $\Delta\costs$ are the bin widths in $W$ and in $\costs$.

The total cross-section $\SI(\GG\to\PP)$ for a given value of 
$\SQS$ is derived from the differential cross-section
${\rm d}\SI(\EE\to\EE\PP)/{\rm d}W$ by using the luminosity
function ${\rm d}\lumi_{\GG}/{\rm d}W$~\cite{Schuler:1996gt}.

The resulting differential cross-sections for the process $\GG\to\PP$ 
in bins of $W$ and $\costs$ are then summed over 
$\costs$ to obtain the total cross-section as a function of $W$
for $\costs<0.6$.

\begin{figure}[htp]
 \centering
  \resizebox{0.4\textwidth}{!}{%
   \includegraphics{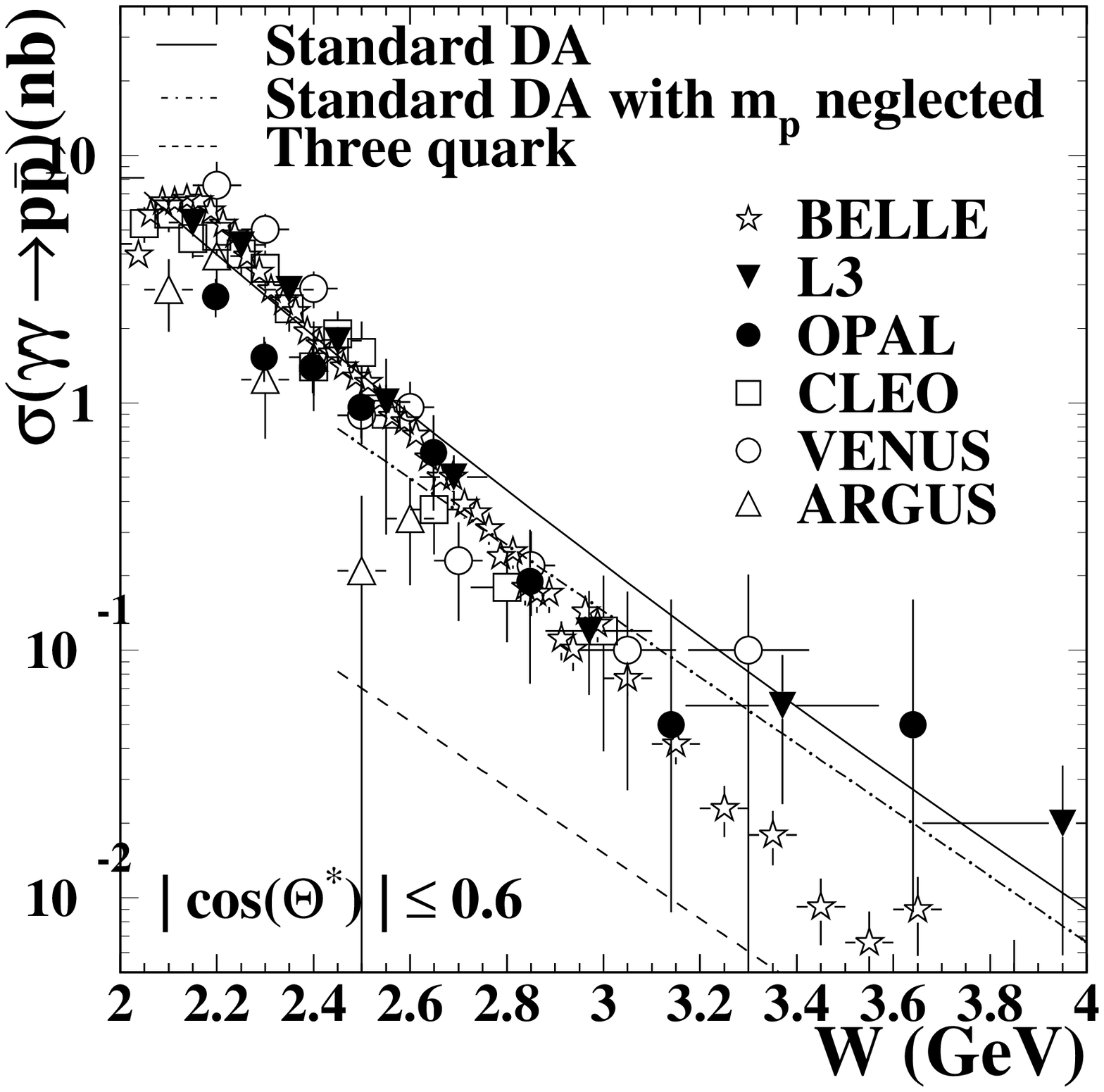}}
  \resizebox{0.4\textwidth}{!}{%
   \includegraphics{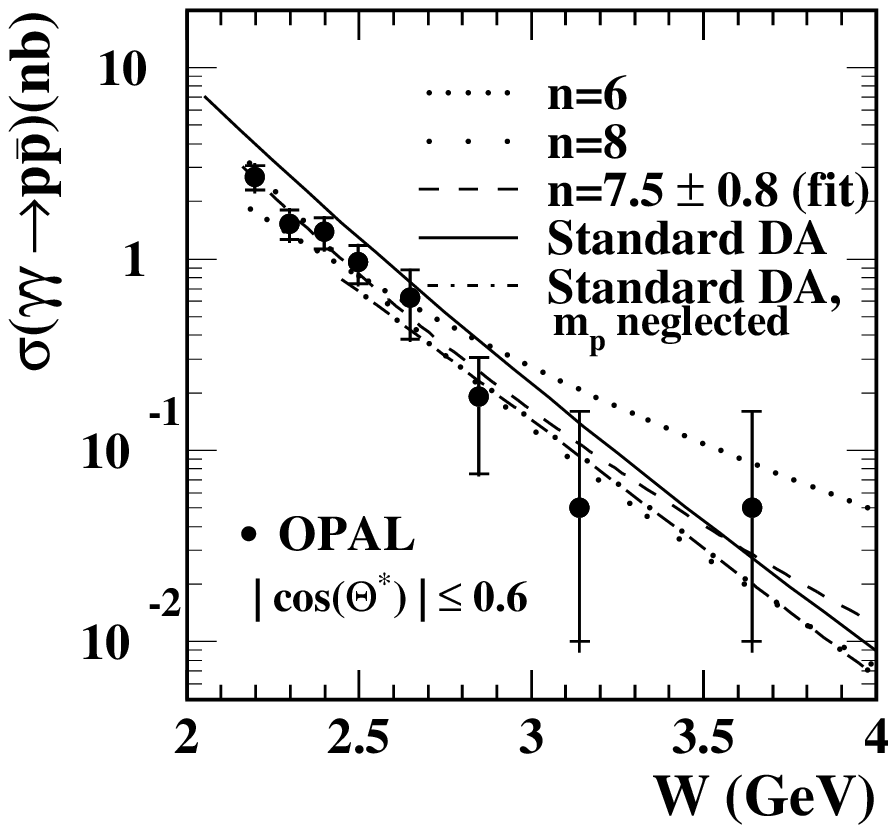}}
   \vspace*{-0.1cm}
    \caption{Cross-sections $\sigma(\GG\to\PP)$ as a function of $W$.
	 The data and the theoretical predictions cover a range of
	 $\costs < 0.6$.  a)(Left plot) The experimental 
data~\protect\cite{Albrecht:1989hz,Artuso:1994xk,Hamasaki:1997cy,Abbiendi:2002bx,Achard:2003jc,Kuo:2005nr}
	 are compared to the quark-diquark model
	 prediction~\protect\cite{berger:2002vc}.  
         The error bars include
	 statistical and systematic uncertainties.  b)(Right plot) The data
	 are compared to the quark-diquark model predictions 
         of~\protect\cite{Anselmino:1989gu}
	 (dash-dotted line), and of~\protect\cite{berger:2002vc} (solid line),
	 using the standard distribution amplitude (DA) with and
	 without neglecting the mass $m_{\rm p}$ of the proton, and
	 with the predictions of the power law with fixed and with
	 fitted exponent $n$.  The inner error bars are the statistical
	 uncertainties and the outer error bars are the total
        uncertainties.}
\label{fig:L3OPALcross}
\vspace*{-0.3cm}
\end{figure}

Fig.~\ref{fig:L3OPALcross}a) shows the cross-section $\SI(\GG\to\PP)$ 
measurements as a function of $W$ for $\costs < 0.6$ obtained 
by ARGUS~\cite{Albrecht:1989hz}, CLEO~\cite{Artuso:1994xk}, 
VENUS~\cite{Hamasaki:1997cy}, OPAL~\cite{Abbiendi:2002bx}, 
L$3$~\cite{Achard:2003jc}, and BELLE~\cite{Kuo:2005nr}. 
Some predictions based on the 
%quark-diquark model~\cite{Ansel:1987vk,berger:2002vc,Kroll:1993zx}, and the 
quark-diquark model~\cite{berger:2002vc,Anselmino:1989gu}, and the 
three-quark model~\cite{Farrar:1985gv} are also shown in this figure.

%\vspace*{-0.3cm}

\noindent There is good agreement between the different experiments results for 
$\WGG > 2.3\GV$. At $\WGG< 2.3\,\GV$ the OPAL~\cite{Abbiendi:2002bx} 
measurements agree with the ARGUS~\cite{Albrecht:1989hz} results, but both 
these measurements lie below the results obtained by CLEO~\cite{Artuso:1994xk}, 
VENUS~\cite{Hamasaki:1997cy}, L$3$~\cite{Achard:2003jc}, and 
BELLE~\cite{Kuo:2005nr}. 

Within the estimated theoretical uncertainties and for $\WGG >2.2\GV$ there 
is a good agreement between the L$3$~\cite{Achard:2003jc} and 
OPAL~\cite{Abbiendi:2002bx} results and the quark-diquark model 
%predictions~\cite{Ansel:1987vk,Kroll:1993zx,berger:2002vc}.
predictions~\cite{berger:2002vc,Anselmino:1989gu}.
The three-quark model is excluded~\cite{Farrar:1985gv}. 
At low $\WGG$ the BELLE~\cite{Kuo:2005nr} results are above 
the quark-diquark model predictions. This measurement agrees with the 
quark-diquark model for $2.5\,\GV < \WGG<3.0\,\GV$, while at higher $\WGG$ a 
steeper fall of the BELLE~\cite{Kuo:2005nr} cross-section is observed.
%
%Nevertheless the BELLE~\cite{Kuo:2005nr} rusults might have  
%underestimated their systematic effects due to e.g. not exclusive 
%$\GG\to\PP$ process background contaminations. 
% 

An important consequence of the pure quark hard scattering 
picture is the power law which follows from the dimensional counting 
rules~\cite{Brodsky:1973kr,Matveev:1973ra}.
%The dimensional counting rules state that an exclusive cross-section 
%at fixed angle has an energy dependence connected with the number of 
%hadronic constituents participating in the process under investigation. 
We expect that for asymptotically large $\WGG$ and fixed 
$\costs$, ${{\rm d}\SI{(\GG\to\PP)}}/{{\rm d}t} \sim W^{2(2-n)}$
%\begin{equation}
%  \frac{{\rm d}\SI{(\GG\to\PP)}}{{\rm d}t} \sim W^{2(2-n)}
%  \label{eq:powerlaw}
%\end{equation}
where $n=8$ is the number of elementary fields
and $t = -W^2/2(1-\costs)$. The introduction
of diquarks modifies the power law by decreasing $n$ to $n=6$.
This power law is compared
to the OPAL data in Fig.~\ref{fig:L3OPALcross}b) with 
$\SI(\GG\to\PP) \sim W^{-2(n-3)}$ using three 
values of the exponent $n$: fixed values $n=8$, $n=6$, 
and the fitted value 
$n=7.5\pm0.8$ obtained by taking into account statistical uncertainties only. 
More data covering a wider range of $W$ would be required to determine the 
exponent $n$ more precisely. 

\begin{figure}[htp]
\begin{center}
\parbox[c]{0.45\textwidth}{%
\vspace*{-0.6cm}
 \centering
  \resizebox{0.42\textwidth}{!}{%
   \includegraphics{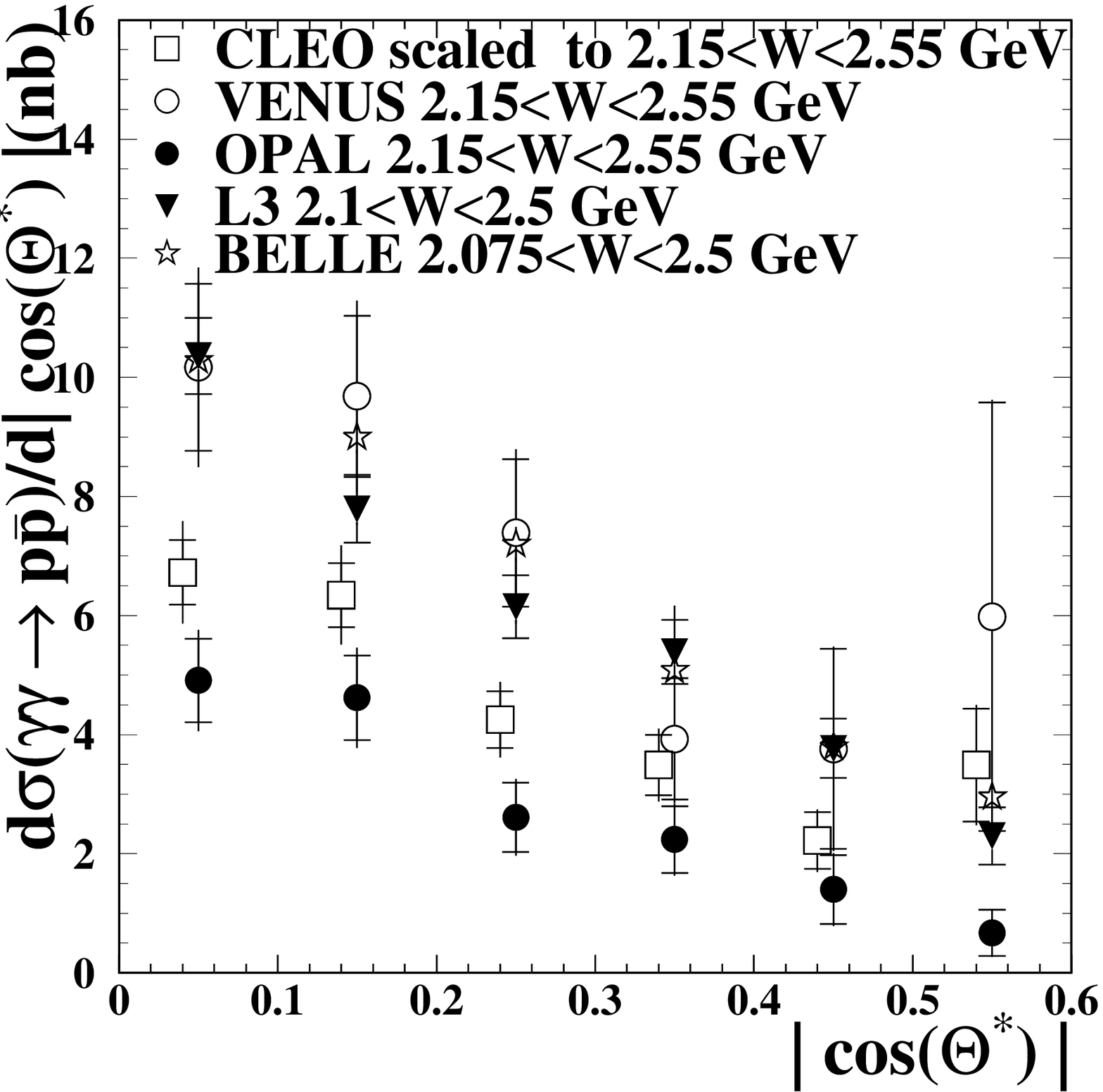}}}
\hspace*{-0.5cm}
\parbox[c]{0.45\textwidth}{%
\vspace*{-0.6cm}
 \centering  
   \rotatebox{270}{\resizebox{0.4\textwidth}{!}{%
   \includegraphics{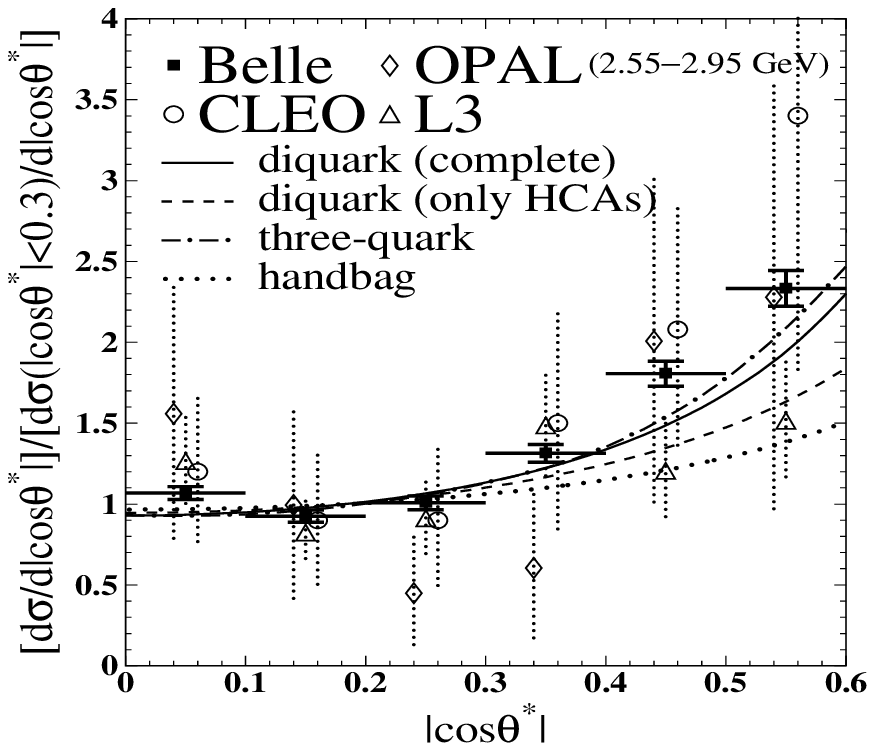}}}}
   \caption{Differential cross-sections for $\GG\to\PP$ as 
            a function of $\costs$ in different ranges of $W$; 
            a) (left plot) low range $2.15<W<2.55\,\GV$, 
            (b) (right plot)
            higher range $2.5<W<3.0\,\GV$.
            The inner error bars are the statistical 
            uncertainties and the outer error bars are the total uncertainties.}
       \vspace*{-0.6cm}
\label{fig:diffcross}
\end{center}
\end{figure}

The measured differential cross-sections 
${{\rm d}\SI{(\GG\to\PP)}}/{{\rm d}\costs}$ in different $W$
ranges and for $\costs<0.6$ 
are shown in Fig.~\ref{fig:diffcross}.

In the range $2.15<W<2.55\,\GV$ the OPAL~\cite{Abbiendi:2002bx} differential 
cross-section lies below the results reported by 
CLEO~\cite{Artuso:1994xk}, VENUS~\cite{Hamasaki:1997cy}, 
L3~\cite{Achard:2003jc}, and BELLE~\cite{Kuo:2005nr}
(Fig.~\ref{fig:diffcross}a)). 
Since the CLEO measurements are given for the lower $W$ range $2.0<W<2.5\,\GV$,
their results have been rescaled by a factor 0.635 which is the ratio of the 
two CLEO total cross-section measurements integrated over the $W$ ranges
$2.0<W<2.5\,\GV$ and $2.15<W<2.55\,\GV$. This leads 
to a better agreement between the OPAL and CLEO measurements but the OPAL 
results are still consistently lower.
The shapes of the $\costs$ dependence of all measurements are 
consistent.

Fig.~\ref{fig:diffcross}b) shows the differential cross-sections 
${{\rm d}\SI{(\GG\to\PP)}}/{{\rm d}\costs}$
in the $W$ range $2.5<W<3.0\,\GV$ obtained by  
CLEO~\cite{Artuso:1994xk}, OPAL~\cite{Abbiendi:2002bx}, 
L3~\cite{Achard:2003jc}, and BELLE~\cite{Kuo:2005nr} in similar $W$ ranges, 
these differential cross-section
have been normalized to the that averaged within $\costs<0.3$.
The measurements are consistent within the uncertainties. 

The comparison of the differential cross-section as a function 
of $\costs$ for $2.55<W<2.95\,\GV$ with the calculation 
of~\cite{berger:2002vc} at $W = 2.8\,\GV$ for different distribution 
amplitudes (DA) is also shown in this figure together with
pure quark model~\cite{Farrar:1985gv} and the 
handbag model prediction~\cite{diehl:2003}.
The shapes of the curves are consistent with those of the data. 
\vspace*{-0.1cm}
Fig.~\ref{fig:diffcross} shows that the differential cross-section at 
low $W$ decreases at large $\costs$, while the opposite trend is observed
in the higher $W$ region. 
The transition point seems to occur at $W\approx 2.5\,\GV$~\cite{Kuo:2005nr}.

\begin{figure}[htbp]
 \centering
  \resizebox{0.4\textwidth}{!}{%
  \includegraphics{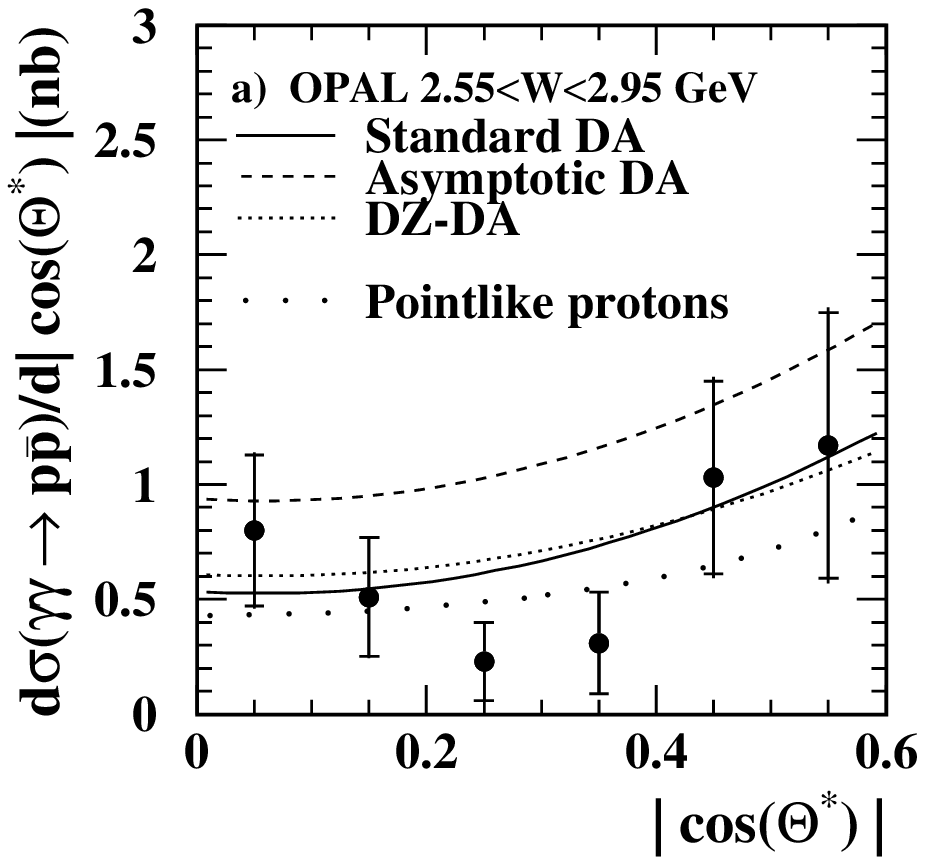}}
  \resizebox{0.4\textwidth}{!}{%
  \includegraphics{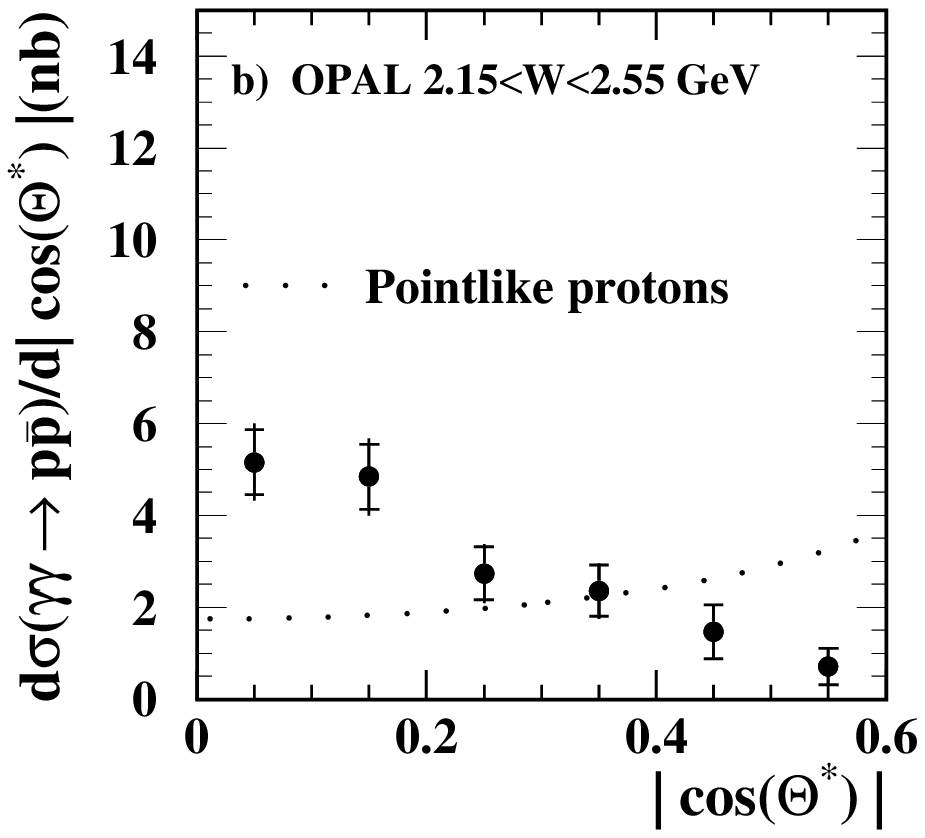}}
  \caption{Measured differential cross-section, 
      ${\rm d}\SI{(\GG\to\PP)}/{\rm d}\costs$, with statistical
      (inner bars) and total uncertainties (outer bars) for 
      a) $2.55<W<2.95\,\GV$ and 
      b) $2.15<W<2.55\,\GV$. 
      The data are compared with
      the point-like approximation for the proton  
      (\ref{eq:costest}) scaled to fit the data.
      The other curves show the 
      pure quark model~\protect\cite{Farrar:1985gv},
      the diquark model of~\protect\cite{Anselmino:1989gu} 
      with
      the Dziembowski distribution amplitudes (DZ-DA), and
      the diquark model of~\protect\cite{berger:2002vc} using standard
      and asymptotic distribution amplitudes. }
  \label{fig:cos2}
\end{figure}

Another important consequence of the hard scattering picture
is the hadron helicity
conservation rule. For each exclusive reaction like
$\GG\to\PP$ the sum of the two initial helicities equals
the sum of the two final ones.
According to the simplification used in~\cite{Anselmino:1989gu}, 
neglecting quark masses, quark and antiquark and hence proton and 
antiproton have to be in opposite helicity states. 
If the (anti) proton is considered as a point-like particle, simple 
QED rules determine the angular dependence of the unpolarized 
$\GG\to\PP$ differential cross-section~\cite{Budnev:1974de}: 
\begin{equation}
  \frac{{\rm d}\SI{(\GG\to\PP)}}{{\rm d}\costs} \propto \frac{(1 + \cos^{2}\theta^{*})}{(1 - \cos^{2}\theta^{*})}.
  \label{eq:costest}
\end{equation}
This expression is compared to the OPAL~\cite{Abbiendi:2002bx} 
data in two $W$ ranges, $2.55<W<2.95\,\GV$ 
(Fig.~\ref{fig:cos2}a) and $2.15<W<2.55\,\GV$ (Fig.~\ref{fig:cos2}b). 
The normalisation in each case is determined by the best fit to 
the data. In the higher $W$ range, the prediction (\ref{eq:costest}) 
is in agreement with 
the data within the experimental uncertainties. 
In the lower $W$ range this 
simple model does not describe the data. At low $W$ soft 
processes such as meson exchange are expected to introduce other partial
waves, so that the approximations leading to (\ref{eq:costest})
become invalid.
\vspace*{-0.6cm}
\vspace*{-0.6cm}
\section{The $\GG\to\LL$ and $\GG\to\SISI$ cross-section measurements}
\label{sec:llbar cross-section}
The cross-sections $\SI(\GG\to\LL)$ and $\SI(\GG\to\SISI)$ in real photon
collisions as a function of $W$ and for $\costs < 0.6$ can be extracted
by deconvoluting the two-photon luminosity function and the form 
factor~\cite{Schuler:1996gt}.

\begin{figure}[htp]
 \centering
  \resizebox{0.3\textwidth}{!}{%
   \includegraphics{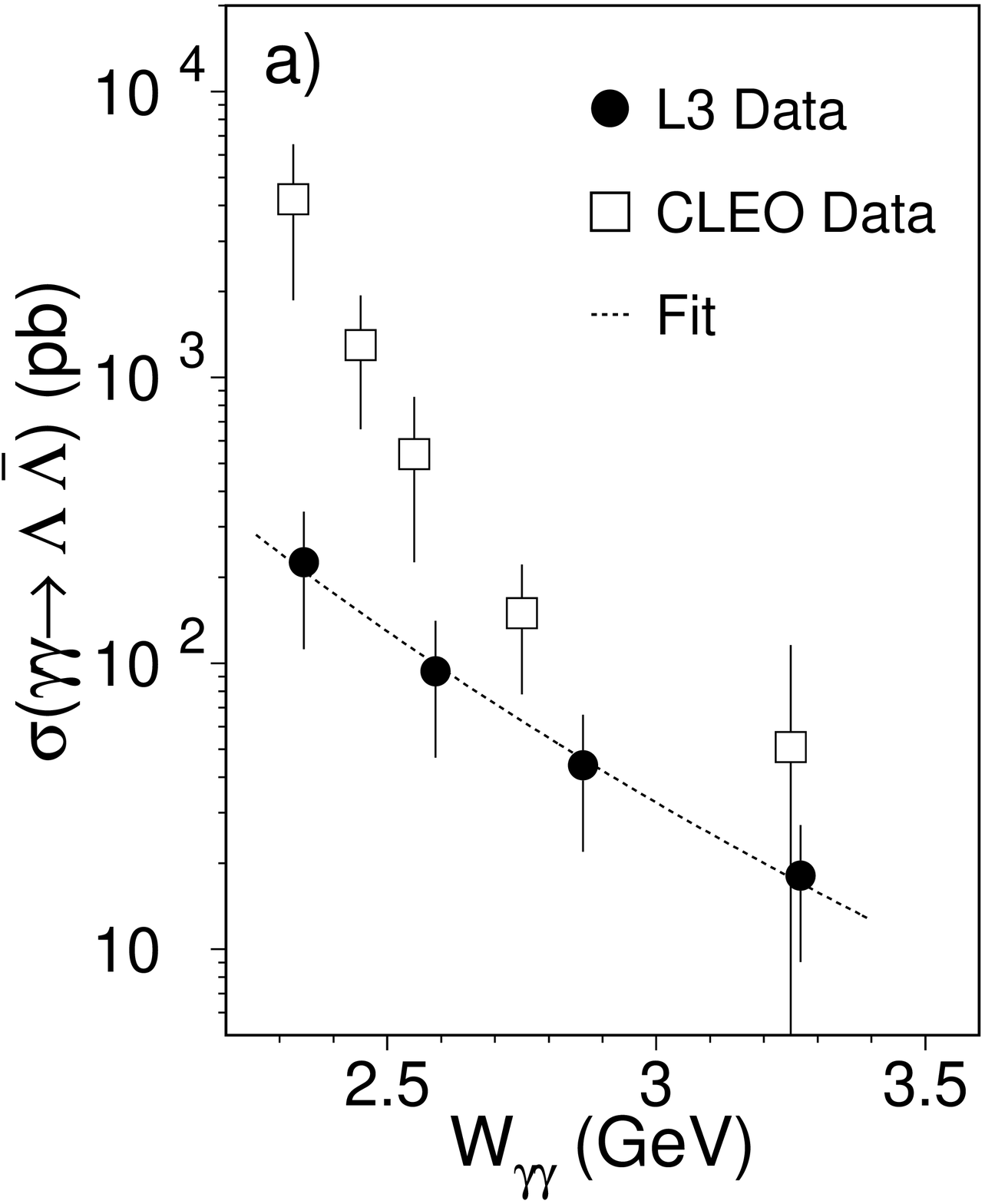}}
  \resizebox{0.3\textwidth}{!}{%
   \includegraphics{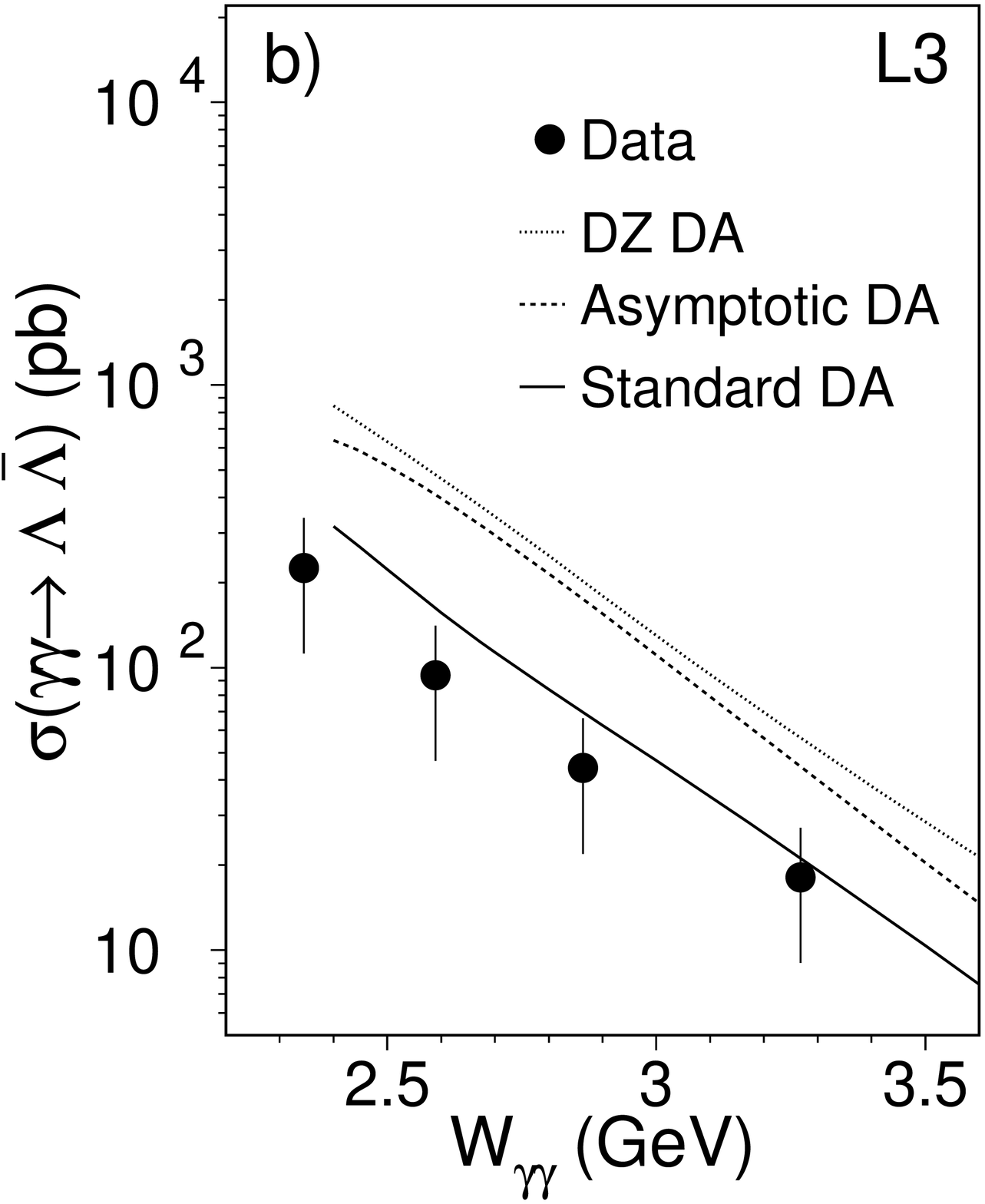}}
  \resizebox{0.3\textwidth}{!}{%
   \includegraphics{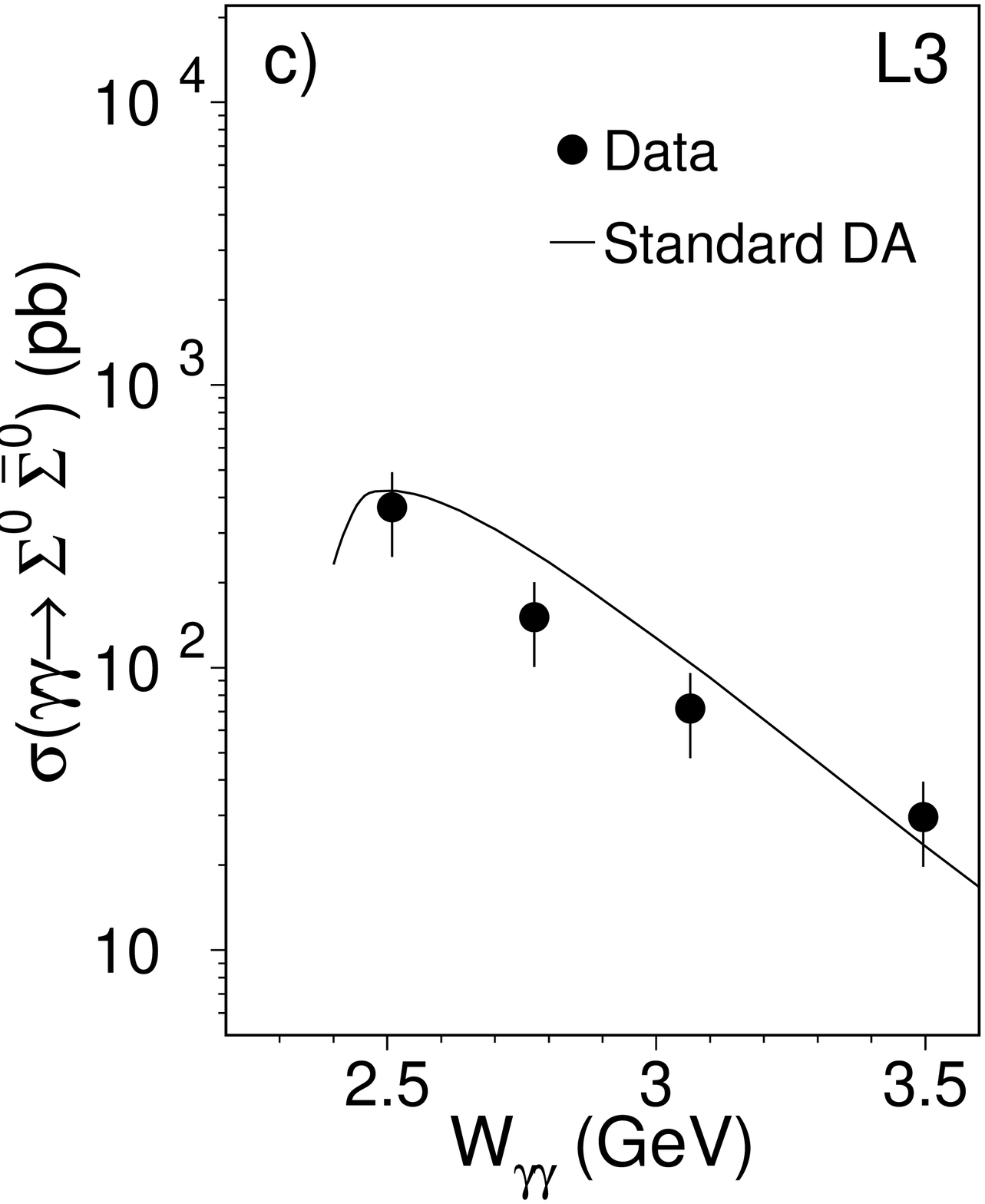}}
    \caption{Measurements of the $\SI(\GG\to\LL)$ and $\SI(\GG\to\SISI)$ 
     cross-sections as a function of $W$. In a) the $\SI(\GG\to\LL)$
     cross-section is compared to the one obtained by 
     CLEO~\protect\cite{Anderson:1997ak}. The dashed line shows the power law fit
     as desccribed in the text. In b) and c) the 
     $\SI(\GG\to\LL)$ and $\SI(\GG\to\SISI)$ measurements are compared
     to the quardk-diquark model predictions~\protect\cite{Berger:2004ye}
      \label{fig:L3Cleo}}
\end{figure}

Fig.~\ref{fig:L3Cleo} compares the L$3$~\cite{Achard:2002ez} 
$\SI(\GG\to\LL)$ measurement with that obtained by 
CLEO~\cite{Anderson:1997ak}.
For $W>2.5\,\GV$ the two results are compatible inside the 
large experimental errors. The cross-section measurement obtained by 
CLEO at lower $W$ values is steeper that the one obtained by L$3$.
The L$3$~\cite{Achard:2002ez} data, fitted with a function of the form
$\SI \approx W^{-n}$, gives a value $n=7.6 \pm 3.9$. 
In Fig.~\ref{fig:L3Cleo} the $\SI(\GG\to\LL)$ and $\SI(\GG\to\SISI)$
cross-section measurements are compared to the predictions of the 
quark-diquark model calculation~\cite{Berger:2004ye}. The absolute predictions 
using the standard distribution amplitude~\cite{Berger:2004ye} (Standard DA) 
reproduce well the L$3$ data, the asymptotic DA and the DZ-DA 
models~\cite{Berger:2004ye} are excluded.
The CLEO~\cite{Anderson:1997ak} and L$3$~\cite{Achard:2002ez} 
$\SI(\GG\to\LL)$ cross-section 
measurements and L$3$ $\SI(\GG\to\SISI)$ cross-section 
measurements for $W>2.5\,\GV$ are satisfactory described also 
by the handbag model, see Ref.~\cite{diehl:2003}.
\vspace*{-0.1cm}

\end{document}